\documentclass[acmsmall]{acmart}

\usepackage{pifont}
\usepackage[most]{tcolorbox}
\usepackage[normalem]{ulem}
\usepackage{threeparttable}
\usepackage{enumitem}
\usepackage{csquotes}
\usepackage{multirow}
\usepackage{url}
\usepackage{hyperref}
\usepackage{fontawesome5} 
\usepackage{algorithm}
\usepackage{algpseudocode}
\usepackage{subcaption}
\usepackage{makecell}

\newcommand{\circled}[1]{\tikz[baseline=(char.base)]{
    \node[shape=circle,draw,inner sep=0.5pt] (char) {#1};}}

\def\eg{\textit{e.g.,} }
\def\ie{\textit{i.e.,} }

\newcommand{\bench}[1]{\textit{ReqEBench}}

\tcbset{
  colframe=black!10,
  colback=white,
  coltitle=black,
  fonttitle=\bfseries,
  boxrule=0.4pt,
  left=6pt,right=6pt,top=6pt,bottom=6pt,
  title style={left color=black!3,right color=black!1}
}

\AtBeginDocument{%
  }
\newtcolorbox{boxK}{
    top=2pt,
    bottom=2pt,
    left=2pt,
    right=2pt,
    boxrule = 0pt,
    toprule = 0pt, 
    colback=gray!15,   
    colframe=white     
}

\AtBeginDocument{%
  }

\setcopyright{acmlicensed}
\copyrightyear{2025}
\acmYear{2025}
\acmDOI{}
\acmJournal{TOSEM}
\acmISBN{}

\begin{document}


\title{Automated Concern Extraction from Textual Requirements of Cyber-Physical Systems: A Multi-solution Study}

\author{Dongming Jin}
\email{dmjin@stu.pku.edu.cn}
\affiliation{%
  \institution{Peking University}
  \city{Beijing}
  \country{China}
}

\author{Zhi Jin}
\email{zhijin@pku.edu.cn}
\authornote{Zhi Jin is Corresponding author}
\affiliation{
  \institution{Peking University}
  \city{Beijing}
  \country{China}
}

\author{Xiaohong Chen}
\email{xhchen@sei.ecnu.edu.cn}
\affiliation{
  \institution{East China Normal University}
  \city{Beijing}
  \country{China}
}

\author{Zheng Fang}
\email{fangz@pku.edu.cn}
\affiliation{
  \institution{Peking University}
  \city{Beijing}
  \country{China}
}

\author{Linyu Li}
\email{xltx_youxiang@qq.com}
\affiliation{
  \institution{Peking University}
  \city{Beijing}
  \country{China}
}

\author{Shengxin Zhao}
\email{orangejuicelike0206@gmail.com}
\affiliation{%
  \institution{Inner Mongolia Normal University}
  \city{Inner Mongolia}
  \country{China}
}

\author{Chunhui Wang}
\email{ciecwch@imnu.edu.cn}
\affiliation{%
  \institution{Inner Mongolia Normal University}
  \city{Inner Mongolia}
  \country{China}
}

\author{Hongbin Xiao}
\email{hongbinxiao@stu.gxnu.edu.cn}
\affiliation{%
  \institution{Guangxi Normal University}
  \city{Guangxi}
  \country{China}
}

\renewcommand{\shortauthors}{Jin et al.}

\begin{abstract}
 
Cyber-physical systems (CPSs) are characterized by a deep integration of the information space and the physical world, which makes the extraction of requirements concerns more challenging. Some automated solutions for requirements concern extraction have been proposed to alleviate the burden on requirements engineers. However, evaluating the effectiveness of these solutions, which relies on fair and comprehensive benchmarks, remains an open question. To address this gap, we propose \bench{}, a new CPSs requirements concern extraction benchmark, which contains 2,721 requirements from 12 real-world CPSs. \bench{} offers four advantages. \ding{182} It aligns with real-world CPSs requirements in multiple dimensions, \eg scale and complexity. \ding{183} It covers comprehensive concerns related to CPSs requirements. \ding{184} It undergoes a rigorous annotation process. \ding{185} It covers multiple application domains of CPSs, \eg aerospace and healthcare. We conducted a comparative study on three types of automated requirements concern extraction solutions and revealed their performance in real-world CPSs using our \bench{}. We found that the highest F1 score of GPT-4 is only 0.24 in entity concern extraction. We further analyze failure cases of popular LLM-based solutions, summarize their shortcomings, and provide ideas for improving their capabilities. We believe \bench{} will facilitate the evaluation and development of automated requirements concern extraction. 
\end{abstract}




\received{20 February 2007}
\received[revised]{12 March 2009}
\received[accepted]{5 June 2009}

\maketitle

\section{Introduction}

\begin{displayquote}
\textsf{``Crosscutting concerns are often scattered and tangled within requirements texts. Their extraction acts as a bridge from unstructured natural language requirements to structured analysis.''}  

\begin{flushright}
\footnotesize ------ Rashid Awais, \textit{RE 2002}~\cite{rashid2002early}; Rago Alejandro, \textit{JSS 2019}~\cite{rago2019concern}.
\end{flushright}
\end{displayquote}

Cyber-physical systems (CPSs) are ubiquitous in modern life~\cite{lee2010cps}, from mobile phones and other electronic products to cars and spacecraft~\cite{menghi2020approximation}~\cite{cornejo2021mutation}. These systems integrate cyber components (\eg software and computational units) with physical devices (\ie sensors, actuators, mechanical devices) and interact with various physical environments continuously~\cite{xu2023semantics}~\cite{mandrioli2023stress}. Thus, they are characterized by the tight coupling of software and physical components. Meanwhile, with the rapid proliferation of heterogeneous sensing and actuating devices, CPSs continue to grow in scale and become increasingly diverse in terms of interactions and applications~\cite{feng2020frepa}. Such trends significantly increase the complexity of CPSs requirements and make the task of understanding and validating requirements texts particularly challenging. 

To address this complexity, \textbf{extracting key concerns from long and unstructured requirements texts} has become a crucial step in the requirements analysis process~\cite{shi2020learning}~\cite{frattini2020automatic}. These concerns typically span multiple types (\ie entity and interaction) and dimensions (\eg physical device and external system), which together form the foundation for downstream modeling and validation tasks. However, the process of extracting these concerns manually is extremely time-consuming and requires a steep learning curve~\cite{zhou2022assisting}. Requirements engineers not only need to carefully read and interpret long documents but also need to possess extensive domain knowledge and have expertise regarding requirements concerns. The steep learning curve and high cognitive burden associated with this task hinder its efficiency. Consequently, despite automated extraction of requirements concerns is highly valuable for easing engineers' workload, its practical adoption in real world CPSs remains limited. This may be because existing solutions lack sufficient adaptability and robustness for complex industrial requirements.

Reviewing previous studies, various artificial intelligence (AI) techniques have been proposed to automatically extract requirements concerns from requirements texts. These solutions fall into three types, \ie heuristic rule-based solutions~\cite{robeer2016automated}~\cite{zaki2022rcm}, machine learning-based solutions~\cite{li2020automated}~\cite{huang2010inducing}, and language model-based solutions~\cite{camara2023assessment}~\cite{ruan2023requirements}. They have shown promising results on their respective benchmarks, but their performance in real-world applications to CPSs is still far from satisfactory~\cite{chen2023automated}. This mismatch suggests that existing benchmarks are poorly aligned with the requirements of real-world CPSs. 
Specifically, the key factor is that existing benchmarks do not adequately reflect the complexity, diversity, and domain specificity of real-world CPS requirements. As a result, they are insufficient for accurately evaluating the practical effectiveness of these approaches, highlighting the need for new benchmarks and methods that can better align with real-world CPS requirements scenarios.

After a systematic and in-depth analysis, we found that the existing benchmarks have three shortcomings:
(1) Insufficient Complexity. The requirements texts in existing benchmarks are mostly made up of hand-crafted cases in textbooks or courses, whose scale and linguistic phenomena (\eg long-range coreference, nested conditionals domain jargon) fall far short of real CPSs. For example, the text of the real-world CPS requirements can span dozens of pages~\cite{feng2020frepa}, but the cases in IT4RE~\cite{al2018use} contain fewer than ten sentences. 
(2) Insufficient Coverage. Existing benchmarks only focus on certain requirements concerns (\eg data function~\cite{li2022automated} and transaction function~\cite{shi2020learning}). They fail to cover the multidimensional concerns of CPSs requirements, including the concerns about multiple individual entities (\eg software system, physical devices, and environment objects) and the critical cross-entity interactions (\eg phenomena linkage and requirements constraint). In addition, application domains are very limited, providing weak evidence of generalization across diverse CPS application domains such as healthcare, aerospace, and smart homes. 
(3) Limited Accessibility. Researchers usually use their own benchmarks for evaluation, but tend not to make them public due to privacy concerns. Worst of all, there are differences and/or inconsistencies between different benchmarks in requirements concern classification and concept annotation. These factors hinder the production of fair comparisons of various solutions.
\textbf{Therefore, in order to better promote the automation of CPSs requirements extraction, it is necessary to construct a public benchmark that can systematically capture the core concerns in real-world CPS requirements.} 

In this paper, we propose \bench{}, a new requirement concern extraction benchmark from CPS requirements, which contains 2,721 requirements from 12 real-world industrial CPSs. \bench{} offers four advantages compared to existing benchmarks. \ding{182} It aligns with real-world CPSs in scale and characteristics of requirements texts. The characteristics include long-range coreference, nested conditionals, and domain-specific jargon. This allows it to effectively evaluate the performance of solutions in the automated extraction of key concerns in real-world CPSs scenarios. \ding{183} It covers more comprehensive concerns related to CPSs requirements. Specifically, \bench{} provides a systematic taxonomy of CPSs requirements concerns. The two top-level categories are related to entities and interactions, respectively, spanning nine core entity concerns (\eg physical device, environment object, and external system), as well as three key interaction concerns that are particularly critical for characterizing CPSs requirements (\eg phenomenon linkage and requirements reference). This multidimensional coverage enables \bench{} to capture not only isolated entity concerns but also critical cross-entity interactions. \ding{184} It undergoes a rigorous annotation process to ensure high quality. Specifically, it has undergone a strict multistage annotation process involving six professional requirements analysts and one CPSs domain expert. A unified annotation guideline was followed and cross-validation was performed to ensure both consistency and high-quality ground truth. This rigorous procedure makes \bench{} a reliable gold standard for evaluating and comparing automated solutions. \ding{185} It spans a wide range of CPSs application domains, including aerospace, healthcare, and smart homes, thus supporting the evaluation of solutions in diverse and heterogeneous contexts. This diversity ensures that the benchmark is not restricted to a single domain but instead facilitates robust assessment of generalization capabilities. Beyond these four advantages, \bench{} is also publicly accessible, enabling the research community to facilitate the evaluation and development of automated CPSs requirements concern extraction from unstructured requirements texts. 

Based on \bench{}, we conduct a comparative study on three types of automated concerns extraction solutions from unstructured requirements texts. The goal of this study is to systematically reveal their strengths and limitations when applied to unstructured requirements texts for real-world CPSs applications. We find that these solutions exhibit low performance on our benchmark. For example, the highest F1 score of GPT-4 is only 0.24 in entity concern extraction. LLM-based solutions significantly outperform traditional methods in concern extraction from requirements texts. For example, GPT-4 with one-shot reasoning achieves an F1 score of 0.30 and a rouge-L of 0.54 on average, while heuristic rule-based and machine learning-based solutions lag far behind. LLM-based solutions can improve their performance with more shots in the prompts. The performance of GPT-4 in the extraction of entity and interaction concerns increased 11\% and 10\% on the F1 score from 1 shot to 2 shots, respectively. LLM-based solutions still do not achieve sufficient performance for practice use, but they have significant potential. The best solution (\ie GPT-4) has a recall of 0.32 on the extraction of entity concerns, indicating that it does not identify more than half of entity concerns. However, it achieves a rouge-L value of 0.55, suggesting that the extraction of entity concerns is usually correct, but there may be occasional vocabulary errors.

We further analyze the failed cases of LLM-based solutions and summarize their error types to provide guidance for future improvements. We observe that the failed cases include four categories: \ie type error, boundary error, omitted error, and complete error. Among them, the boundary error is the most common type and is responsible for 35\% of all failed cases. This is because LLM-based solutions have insufficient understanding of domain-specific terminology. We also observe that the current solutions all perform poorly in extracting the \textit{requirements} concerns\footnote{\textit{requirements} is an entity concept in our taxonomy. See Section~\ref{subsec:taxonomy}}. For example, LLM-based solutions only achieve about 0.1 in F1 in \textit{requirements} entity concern, but they can achieve about 0.3 in other types of entity-level concerns. This is because \textit{requirements} entity concerns often involve abstract and context-dependent descriptions, making it difficult for LLMs to capture their semantic boundaries accurately.

The experimental results reveal the need for an in-depth exploration of the following two aspects. One is designing retrieval-augmented strategies or instruction tuning to inject the definitions of various CPSs requirements concerns into LLMs, which may reduce type errors, omission errors, and complete errors. The other is using continued pre-training with CPS-specific corpora (\eg communication protocols) to enhance the extraction of domain-specific terminology and long-tail requirements, thereby mitigating the most common error (\ie boundary error) and improving the extraction of entity concerns. In addition, future work should focus on improving the modeling of interaction concerns, which require understanding the relational semantics between multiple entities (\eg device-to-system). Incorporating structured reasoning or relation-aware graph representations may help LLMs capture these cross-entity dependencies more effectively. We hope that our \bench{} can facilitate the evaluation of such emerging techniques and further promote the development of automated requirements concern extraction in the future.

We summarize our contributions as follows.
\begin{itemize}
    \item We perform a systematic analysis for evaluation data on requirements concern extraction from three aspects, \ie text types, text sources and focused concerns. 
    \item We present a comprehensive characterization of key concerns in CPSs requirements, serving as a foundation for systematic extraction and evaluation.  
    \item We propose \bench{}, a new benchmark for the extraction of requirements concerns from CPS requirements texts, which bridges the gap between evaluation and application. 
    \item We develop three types of automated concerns extraction solutions and conduct a comparative study on them, revealing their weaknesses in dealing with CPSs requirements. 
    \item We analyze the failed cases of advanced LLM-based solutions to summarize the shortcomings of existing solutions and provide ideas for future improvement.     
\end{itemize}

\textbf{Data Availability}. We open source our replication package~\cite{web:code}, which includes the benchmark \bench{} and the evaluation source code, to allow other researchers and practitioners to replicate our work and validate their studies.

\section{Analysis on Requirements Concern Extraction and Motivation}

This section introduces a small-scale systematic literature review on the extraction of requirements concerns from the evaluation dataset perspective and describes the motivation for constructing a new benchmark for the extraction of CPSs requirements concerns. 

\subsection{A Systematic Analysis on Requirements Concern Extraction from the Evaluation Data Perspective}~\label{sec:survey}

\textbf{Study Goals.} The goal of this analysis is to investigate how existing studies on the requirements concern extraction have been evaluated with a particular focus on the datasets used. We will answer the following questions(Q). 

\begin{table}[]
    \centering
    \caption{Number of papers filtered in each iteration.}
    \begin{tabular}{lcccc}
\toprule
               & Iter 1 & Iter 2 & Iter 3 & Iter 4 \\ \midrule
\textbf{Total} & 230    & 98       & 47     & 21        \\ \bottomrule
\end{tabular}
    \label{tab:selectedpapers}
\end{table}

\textbf{Q1: What types of evaluation data are employed in previous studies?} The type of evaluation data plays a crucial role in assessing the effectiveness of solutions, which directly affects the robustness and generalizability of results. The types of evaluation data can be classified into two categories: case studies, which rely on a small number of requirements texts to demonstrate feasibility, and evaluation datasets, which consist of a larger collection of requirements texts. To obtain a clear picture, we analyze the type of evaluation data adopted in each study on requirements concern extraction and report their distribution accordingly.   

\textbf{Q2: What are the sources of requirements texts in their datasets?} The source of requirements texts directly influences the complexity and distribution of the data, thereby shaping the validity of the evaluation results. These sources can be classified into three categories: \ie real-world industrial requirements texts, tutorial materials (\eg textbooks or course assignments), and synthesized datasets. Although tutorial and synthesized data are easier to obtain and annotate, they fail to capture many phenomena in real CPSs requirements. To clarify this aspect, we examine the source of requirement texts used in previous studies and summarize their proportions separately. 

\textbf{Q3: What kinds of concerns are typically targeted for extraction in their datasets?} The scope of concerns selected for extraction in datasets also has a direct impact on the evaluation outcomes. Most existing studies focus on a narrow set of concerns, \eg data functions, transaction functions. While this reduces annotation complexity, it overlooks broader and more representative concerns in CPSs requirements, including entities (\eg physical devices, environment objects) and cross-entity interactions (\eg interfaces). To better understand current coverage, we analyze the types of concerns focused in previous studies. 

\textbf{Collected Papers.} To identify relevant papers for our analysis, we adopted a systematic search strategy following previous work~\cite{zhang2023survey}. Initially, we determine keywords derived from our formulated questions, \ie \textit{extract} and \textit{extraction}. Since requirements texts are often described under various terminologies (\eg user story and app review), and concerns are usually expressed as concrete categories (\eg data functions, transaction functions) rather than explicitly using the term ``concerns'', we deliberately avoided overly specific keywords to ensure greater coverage. This approach inevitably searched for many irrelevant papers, but these were subsequently removed during the screening process. Our search spanned 7 journals\footnote{The journals comprised TOSEM, TSE, ASE, ESE, IST, JSS, and REJ} and 9 conferences\footnote{The conferences included ICSE, FSE, ASE, ISSTA, RE, CAiSE, MoDELS, SANER, and ICSME} in the software engineering community, covering papers published from January 1, 2015 to September 27, 2025. The initial search for these journals and conferences yielded a total of 230 papers. 

To further ensure the quality of these selected papers and their relevance, we use the following factors to include or exclude a paper, beyond our filtering process. Our criteria for including a work as relevant in the analysis were the following:
\begin{itemize}
    \item The paper must explicitly address the task of extracting information from requirements texts, such as requirements documents, user stories, or app reviews.
    \item The paper must utilize natural language processing techniques to propose solutions for automating or assisting in the concern extraction from the requirements texts. 
    \item The paper must provide details on the evaluation dataset and report performance results.
\end{itemize}
We also established a set of exclusion criteria which are adopted from previous research as follows:
\begin{itemize}
    \item Short papers whose number of pages is less than or equal to 4.
    \item Duplicate papers or similar studies with different versions by the same authors.
\end{itemize}

Then we conducted a four-iteration approach to ensure a comprehensive and unbiased selection of relevant papers, adapted from ~\cite{rodriguez2021perceived}. The screening process is as follows:
\begin{itemize}
    \item \textbf{Title Screening:} The first author reviewed the titles of all 230 papers, categorizing them as \textit{Include} or \textit{Exclude} based on our predefined criteria mentioned above and the scope of the study. 
    \item \textbf{Abstract Screening:} For papers that passed title screening, the first author examined the abstracts, again applying \textit{Include} and \textit{Exclude} categorization. Papers labeled as \textit{Uncertain} were investigated more carefully in the next step.
    \item \textbf{Preliminary Content Review:} The first author conducted a cursory examination of the full text of the papers that were marked as \textit{Include} in the previous iteration, to further assess their relevance.
\end{itemize} 

Table~\ref{tab:selectedpapers} details the number of papers we initially found and after each of the filtering steps. In total, of the 230 papers, we filtered 21 papers to read in the final list. We read all the papers in this list to extract information about the evaluation data and answer each of the questions. 

\begin{table}[]
    \centering
    \caption{The answers to all three questions in our systematic analysis}
    \begin{tabular}{cccc}
\toprule
\textbf{Papers}        & \textbf{Answer to Q1} & \textbf{Answer to Q2} & \textbf{Answer to Q3} \\ \midrule
Arora et al, TSE, 2016~\cite{arora2016automated} & Case Study (3)    &  Industrial  &  Glossary Terms        \\
Robeer et al, RE, 2016~\cite{robeer2016automated} & Case Study (2)   & Industrial   & Conceptual Models      \\
Arora et al, MoDELs, 2016~\cite{arora2016extracting} & Case Study (4)  & Industrial  &  UML Models          \\
AlHroob et al, IST, 2018~\cite{al2018use}     & Case Study (5)    & Tutorial    &  Action, Actor         \\
Gemkow et al, RE, 2018~\cite{gemkow2018automatic}  & Dataset   & Industrial    & Glossary Terms      \\
Arora et al, TOSEM, 2019~\cite{arora2019active}  & Case Study (3)  & Industrial  & UML Models      \\
Pudlitz et al, RE, 2019~\cite{pudlitz2019extraction} & Dataset  & Industrial & System States      \\
Frattini et al, ASE, 2020~\cite{frattini2020automatic} & Dataset & Industrial & Cause-Effect      \\
Gunecs et al, RE, 2010~\cite{gunecs2020automated} & Dataset  & Tutorial & Goal Model             \\
Fischbach et al, RE, 2020~\cite{fischbach2020towards} & Dataset & Tutorial & Cause-Effect        \\
Li et al, ICSME, 2020~\cite{li2020automated} & Dataset & Industrial & Entities                       \\
Javed et al, IST, 2021~\cite{javed2021imer} & Case Study (3) & Tutorial& Process Model \\
Zaki et al, ASEJ, 2022~\cite{zaki2022rcm} & Dataset & Synthesis& Action, et al \\
Li et al, IST, 2022~\cite{li2022automated} & Dataset & Industrial & Data Function     \\
Yang et al, MoDELS, 2022~\cite{yang2022towards} & Dataset & Tutorial & UML Models      \\
Fischbach et al, JSS, 2023~\cite{fischbach2023automatic} & Case Study (1) & Industrial & Test Condition \\
Arulmohan et al, MoDELS, 2023~\cite{arulmohan2023extracting} & Dataset & Tutorial & Person, et al. \\
Lim et al, JSS, 2024~\cite{lim2024test} & Case Study (3) & Industrial  & Actor, et al.  \\
Das et al, JSS, 2024~\cite{das2024extracting} & Case Study (2) & Industrial & Goal Model        \\
Siddeshwar et al, MoDELS, 2024~\cite{siddeshwar2024comparative} & Case Study (10) & Tutorial & Goal Model                       \\
Zhang et al, CAiSE, 2025~\cite{zhang2025litroacp} & Dataset & Industrial & Access Control Policy    \\ 


\bottomrule
\end{tabular}
    \label{tab:analysisresult}
\end{table}

\textbf{Results and Analysis.} Table~\ref{tab:analysisresult} summarizes the results of our systematic analysis of the 21 selected papers. We report the answers to the three guiding questions for each paper. We can observe that nearly half of the papers (10/21) rely on small-scale case studies for evaluation. This trend highlights a lack of scalability in evaluation and suggests that many solutions have only been validated in controlled settings. We also found that industrial requirements texts are the most frequently adopted (12/21), reflecting the importance of real-world complexity for validating extraction approaches. In addition, most studies focus on narrow categories (\eg data function and transaction function) or a certain requirements modeling model (\eg UML model and goal model). This indicates that existing datasets do not fully represent the multidimensional nature of CPSs requirements, which typically involve entities (\eg physical devices, environment objects) and interactions (\eg phenomena linkage, requirements constraint).

 \begin{table}[]
     \centering
     \setlength{\tabcolsep}{4pt}
     \caption{Comparison between existing benchmarks and \bench{}}
     \begin{tabular}{lcccccc}
\toprule
\textbf{Existing benchmark}         & \textbf{Real}      & \textbf{Com}  & \textbf{Inter} & \textbf{Annota}      & \textbf{Multi}          & \textbf{Pub} \\ \midrule
Aysh et al, 2018~\cite{al2018use}         & \faTimes         & \faTimes                      & \faTimes               & \faTimes               & \faCheck               & \faCheck             \\
Javed et al, 2021~\cite{javed2021imer}          & \faTimes         & \faTimes                      & \faTimes               & \faTimes               & \faCheck               & \faCheck             \\
Aya et al, 2022~\cite{zaki2022rcm}            & \faTimes         & \faTimes                      & \faTimes               & \faTimes               & \faCheck               & \faCheck             \\
Zhou et al, 2022~\cite{zhou2022assisting}           & \faCheck     & \faTimes                      & \faTimes               & \faCheck           & \faCheck               & \faCheck             \\
Camera et al, 2023~\cite{camara2023assessment}         & \faTimes         & \faTimes                      & \faTimes               & \faCheck           & \faCheck               & \faTimes              \\
Jin et al, 2023~\cite{jin2023automating}            & \faCheck     & \faTimes                      & \faCheck               & \faTimes               & \faTimes                   & \faTimes              \\
Ruan et al, 2023~\cite{ruan2023requirements}           & \faCheck     & \faTimes                      & \faTimes               & \faCheck           & \faTimes                   & \faCheck             \\
Chen et al, 2023~\cite{chen2023automated}           & \faTimes         & \faTimes                      & \faTimes               & \faCheck           & \faCheck               & \faTimes              \\

\midrule
\textbf{\bench{} (Ours)} & \faCheck     & \faCheck                  & \faCheck           & \faCheck           & \faCheck               & \faCheck             \\ \bottomrule
\end{tabular}
     \label{tab:feature_compare}
 \end{table}
 
\subsection{Motivation}
After a systematic and in-depth analysis in Section~\ref{sec:survey}, we then introduce our motivation to construct a new benchmark for CPSs requirements concern extraction and conduct a comparative study with previous solutions. 

\textbf{The performance on existing benchmarks does not align with the actual experiences of practitioners in real-world CPSs.} Existing benchmarks consist mainly of small hand-crafted cases from textbooks or courses. They are not fully aligned with real-world CPSs, which creates a gap between evaluation and practical applications. To accurately evaluate the effectiveness of requirements concern extraction, it is necessary to construct a new benchmark that aligns with the real-world CPSs requirements.

To achieve this goal, we first analyze more than 20 CPSs from an aerospace technology company and conclude that an effective benchmark of requirements concern extraction should exhibit the following features. 

\begin{itemize}
    \item \textbf{\textit{Real}-world CPSs}. 
    The benchmark should be based on real-world CPSs requirements to ensure that its evaluation results are consistent with the analyst's actual experience.
    \item \textbf{\textit{Com}prehensive Perspective}. 
    The concerns that the benchmark covers should reflect the multidimensional characteristics of CPSs to ensure that the CPSs requirements can be fully expressed.
    \item \textbf{Rigorous \textit{Annota}tion}. 
    The benchmark should have a clear annotation guideline and qualified experts should be engaged for annotation and cross-validation to ensure its high quality.
    \item \textbf{\textit{Multi}ple domains}. The benchmark should include CPSs requirements from different domains to ensure that it does not have biases towards any single domain or system. 
    \item \textbf{\textit{Pub}lic Access}. The benchmark should be publicly accessible to allow researchers to evaluate and compare their own solutions and tools.
\end{itemize}

We evaluated 8 existing benchmarks using the above criteria and found that none of them can meet all features, as shown in Table \ref{tab:feature_compare}. This gap affects the usability assessment and the improvement of CPSs requirements concern extraction. Therefore, we propose a new benchmark of requirements concern extraction that can satisfy the above features. 

\textbf{Lack of fair comparison on previous solutions.} Most solutions to automate the extraction of requirements concerns have been evaluated using their own hand-packed cases or datasets, which are customized to highlight the strengths of the respective methods. The reliance on such limited, hand-picked cases may distort the performance and fail to reflect actual performance in practice. In addition, the lack of consistent evaluation across various solutions makes it difficult for practitioners to find the most effective solution. Therefore, we conduct a comparative study using a unified benchmark (\ie \bench{}) to highlight the strengths and weaknesses of various extraction solutions.

\section{Benchmark Construction} \label{sec:benchmark}
\begin{figure*}
    \centering
    \includegraphics[width=\linewidth]{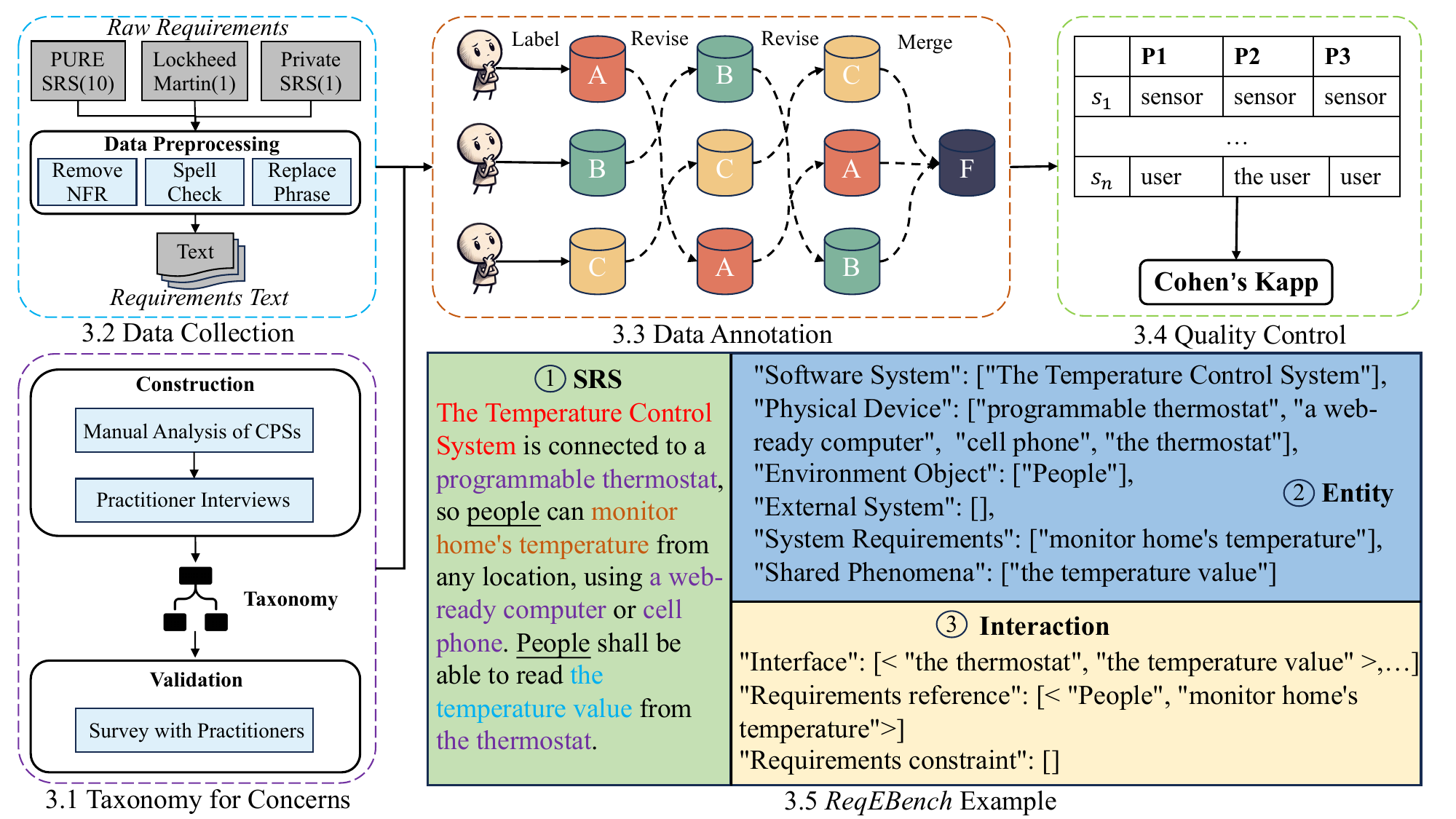}
    \caption{The Overview of \bench{} Construction}
    \label{fig:pdd}
\end{figure*}

In this section, we propose a CPSs \textbf{Req}uirements Concern \textbf{E}xtraction \textbf{Bench}mark named \bench{}. Figure \ref{fig:pdd} illustrates the process of constructing the benchmark. The following subsections present the details of the process, including three steps and an example case from the benchmark. 

\subsection{Taxonomy for CPSs requirements Concerns}~\label{subsec:taxonomy}
To ensure that \bench{} covers comprehensive key concerns in CPSs requirements, we first establish a taxonomy for CPSs requirements concerns. 

\textbf{Construction.} To construct a taxonomy that systematically captures CPSs requirements concerns, we ran a mixed-methods approach comprising a series of studies: a literature-based analysis and an interview-based study. \textbf{(1) Literature-based Analysis:} We first conducted a focused literature review to identify the fundamental characteristics of CPSs that differentiate them from traditional software systems. According to Lee~\cite{lee2008cyber} and Rajkumar et al.~\cite {rajkumar2010cyber}, CPSs are deeply integrated systems that combine computation, networking, and physical processes, in which embedded computers and networks monitor and control physical entities through continuous feedback loops. These systems are characterized by tight coupling between cyber and physical components, safety- and time-critical operation, and continuous interaction with dynamic environments~\cite{baheti2011cyber,derler2012modeling}.
From these defining properties, we identified two fundamental dimensions that can shape CPSs requirements:
the \textit{entity} dimension, representing the software, hardware, and environmental components that constitute the system; and the \textit{interaction} dimension, representing the dependencies, control flow, and safety constraints at an interface of shared phenomena (e.g. events, states, values, etc.) between software and other entities.
This theoretical grounding ensures that our taxonomy reflects the intrinsic structure of CPSs and not merely an ad-hoc categorization of requirements.\textbf{(2) Interview Study:} To complement the literature perspective with practical insights, we conducted structured interviews with three CPSs domain experts and practitioners from industrial and research backgrounds. Each participant had at least one year of professional experience in CPS requirements analysis.
The interviews followed a semi-structured format, focusing on the key concerns that practitioners typically address when analyzing CPS requirements (\eg safety, interoperability, real-time performance, and environmental interactions). The experts’ responses helped validate whether the concerns derived from literature indeed capture the issues encountered in real-world CPS projects and also introduced new, practitioner-grounded aspects—particularly related to the operational environment and hardware–software coupling.

\textbf{Formation.} Integrating the theoretical foundations from the literature and the practical evidence from the interviews, we established a taxonomy comprising two top-level categories—\textit{entity} and \textit{interaction}. The entity refers to the fundamental objects given in CPS requirements texts and includes 5 specific concerns. Below is a brief introduction to the entity concerns.

\begin{itemize}
    \item Software System (\textbf{SS}): the software system to be developed, \eg \textit{the smart home control system}.
    \item Physical Device (\textbf{PD}): the real-world physical devices, which can be used to send or receive data, \eg \textit{the sensor} and \textit{the air conditioner}.
    \item Environment Object (\textbf{EO}): the external objects in the interactive environment, \eg \textit{the user} and \textit{the operator}.
    \item External System (\textbf{ES}): the third-party systems that already exist with predefined properties. Their properties are artificially designed or prescribed, \eg \textit{database}.
    \item System Requirements (\textbf{SR}): the purpose of a software system, \eg \textit{control the home environment}.
    \item Shared Phenomena (\textbf{SP}): the observable events that happen between between software and other entities in the system (\eg \textit{dataflow, state, operation}). 
\end{itemize}

The interaction refers to the relationships and dependencies between these entities and includes 3 specific concerns. Below is a brief introduction to these interaction concerns. 

\begin{itemize}
    \item Phenomena Linkage (\textbf{PL}): describes how a shared phenomenon connects different parts of the system, \eg \textit{(the smart home control system, the notification)}. It captures the link between a phenomenon and the entities involved in it.
    \item Requirements Linkage (\textbf{RL}): represents when a requirement explicitly mentions or depends on another entity (\ie PD, EO, and ES), \eg \textit{(the patient, monitor the health condition)}. This linkage shows how requirements are tied to the entities that make them possible.
    \item Requirements Constraint (\textbf{RC}): refers to the rules or limits that a requirement imposes on other parts of the system (\ie PD, EO and ES), \eg \textit{(the medical watch, to monitor patient)}. This type of linkage emphasizes that requirements not only describe behavior but also restrict how entities should operate.
\end{itemize}

\begin{table}[]
    \centering
    \caption{Validation Survey Results for the CPS requirements Concerns}
    \begin{tabular}{lcccccccc}
\toprule
\multicolumn{1}{c}{\multirow{2}{*}{\textbf{Concern}}} & \multicolumn{2}{c}{\textbf{Response}} & \multicolumn{3}{c}{\textbf{Importance}} & \multicolumn{3}{c}{\textbf{Effort Required}} \\
\multicolumn{1}{c}{}                                  & Yes                & No               & Minor      & Major      & Critical      & Low          & Medium         & High         \\ \midrule
Software System                                       &                    5&                  0&            0&            1&               4&              0&                2&              3\\
Physical Device                                       &                    5&                  0&            0&            2&               3&              0&                1&              4\\
Environment Object                                    &                    4&                  1&            0&            3&               2&              0&                1&              4\\
External System                                       &                    5&                  0&            0&            1&               4&              0&                2&              3\\
System Requirements                                   &                    5&                  0&            0&            0&               5&              0&                1&              4\\
Shared Phenomena                                      &                    4&                  1&            0&            2&               3&              0&                2&              3\\ \midrule
Phenomena Linkage                                             &                    5&                  0&            0&            1&               4&              0&                2&              3\\
Requirements Linkage                                &                    5&                  0&            0&            1&               4&              1&                1&              3\\
Requirements Constraint                               &                    4&                  1&            1&            1&               3&              1&                1&              3\\ \bottomrule
\end{tabular}
    \label{tab:taxo_survey}
\end{table}

\textbf{Validation.} To ensure that the final taxonomy is comprehensive and representative of real-world CPS requirements concerns, we validated it through a survey involving a new set of practitioners/researchers, different from those who participated in the construction. We recruit practitioners for the survey from the authors' personal contacts, and 5 participants agreed to take part in our survey. They are all CPS requirements engineers, with a minimum overall requirements analysis experience of 1 year and a maximum of 4 years. We created our survey using Qualtrics~\cite{qualtrics}, a web-based tool to conduct survey research, evaluations, and other data collection activities. In the survey form, we presented the name of each concern, its textual description, and then three questions associated with it. The first question was a ``yes'' or ``no'' question about whether the participant agreed to this concern. In case of a positive answer, we had two more Likert-scale questions on the importance of the concern and the amount of effort required to extract it. In this way, we evaluated not only the acceptance of the taxonomy but also the perceived importance and practical effort of each concern by practitioners. 
The results of the validation survey are summarized in Table~\ref{tab:taxo_survey}. We can observe that all the concerns in our taxonomy are acknowledged by at least four out of five participants, confirming their practical relevance. The most approved concern is \textit{system requirements}, with unanimous ``yes'' answers (100\%). This concern also shows the highest importance, with all participants rating it critical and 80\% indicating high effort for extraction. \textit{Physical device} and \textit{external system} also reached full approval (100\%), with a majority of critical ratings and high effort assessments. The least approved concerns are \textit{environment object}, \textit{shared phenomena}, and \textit{requirements constraint}, which are also confirmed by 80\% of the participants and still have a substantial fraction. On average, participants rated 8.9 out of 9 concerns as relevant, and the distributions in all concerns lean strongly toward critical importance and high effort. These results confirm that the taxonomy captures concerns that are both widely recognized and practically demanding in CPSs requirements analysis. 

\begin{table}[]
    \centering
    \setlength{\tabcolsep}{5pt}
    \caption{The Collected CPS requirements Documents}
    \begin{tabular}{lllll}
\toprule
\textbf{CPSs}   & \textbf{Notion}         & \textbf{Domain}        & \textbf{Source}          & \textbf{Page} \\ \midrule
Crime Tracking Systems & CTS  & Security      & PURE            & 19   \\
Mars Mission System  & MMS & Aerospace     & PURE            & 17     \\
Space Fraction System & SFS               & Aerospace     & PURE            & 9    \\
Tactical Control System  & TCS            & Military      & PURE            & 148  \\
Correlator Control System & CCS  & Signal        & PURE            & 17   \\
Home Control System  & HCS         & Healthcare    & PURE            & 15   \\
Gemini Control System & GCS              & Aerospace     & PURE            & 96   \\
Lane Control System & LCS      & Transportation & PURE            & 66     \\
Flight Control System  & FCS     & Aerospace     & PURE            & 107  \\
Center-to-Center System & C2C     & Networks      & PURE            & 32   \\ \midrule
Autopilot Control System & ACS            & Aerospace     & LM & 4      \\ \midrule
Sun Search Control System & SSCS         & Aerospace     & Industry        & 38     \\ \bottomrule
\end{tabular}
    \label{tab:srs}
\end{table}

\subsection{Data Collection}

\textbf{Studied systems.} This study aims to construct a public benchmark of requirements concern extraction aligned with real-world CPSs. To achieve this, we collect CPSs requirements texts covering diverse application domains, including embedded systems, control systems, real-time systems, etc. The texts originate mainly from two sources: datasets of public software requirements documents (\ie PURE~\cite{ferrari2017pure} and Lockheed Martin (LM)~\cite{mavridou2020ten}), and private requirements documents from an aerospace industry~\cite{yang2022intelligent}. 

PURE consists of 79 documents on software requirements from various domains. We manually reviewed these documents and selected CPSs documents from them. Lockheed Martin provides 10 CPS requirements documents. We also reviewed these documents and selected an autopilot control system. The criteria for selecting this document are the project scale and the areas of applicability. Specifically, the remaining nine requirements documents are approximately one page long. For our private requirements, we selected a Sun search control system (\ie SSCS) from an aerospace company. The SSCS has been anonymized to remove technical details while preserving its functional semantics. It has been used in previous studies for evaluation~\cite{chatmodeler}~\cite{projection}. In total, the collected CPS requirements documents are summarized in Table~\ref{tab:srs}.

\textbf{Requirements texts pre-processing}. The raw software requirements documents often contain tables, figures, and incomplete sentences. To ensure the quality of the requirements documents, we clean and sample the raw documents from the following four aspects: (1) Remove Irrelevant Content: we first remove the catalog, titles, diagrams, and tables using the regex tool and PyPDF2 library. We also remove sentences with no more than 10 words using the regex tool, as these sentences tend to be noisy data that do not contain valid content. (2) Spell Check: we perform a thorough spell check on the remaining requirements descriptions to ensure that any typographical errors are corrected. We also rewrite incomplete sentences by hand because these sentences may create ambiguity, and here we focus on disambiguation for requirements modeling, rather than requirements disambiguation in general. (3) Replace Phrase: We replace specific phrases and terminologies that are either ambiguous or inconsistent with standardized terminologies. For example, we replace terms (\eg \textit{TCS}) with more specific phrases (\eg \textit{Tactical Control System}). (4) We split requirements documents into sentences using Spacy tools~\cite{vasiliev2020natural}. It should be noted that we made only minimal edits to the original requirements, focusing primarily on correcting spelling errors and addressing terminology inconsistencies, which account for approximately 3-5\% of the documents. These edits are made to ensure clarity and consistency, rather than to simplify the task or modify the content of the requirements documents.

\begin{table}[]
    \centering
    \caption{The statistics of the \bench{} benchmark.}
    \begin{tabular}{cccccccccc}
\toprule
\multirow{2}{*}{\textbf{CPSs}} & \multicolumn{6}{c}{\textbf{Entities}} & \multicolumn{3}{c}{\textbf{Interactions}} \\ \cmidrule(lr){2-10}
                     & SS  & PD & EO & ES & SR & SP & PL       & RL       & RC      \\ \midrule
CTS+                 & 100 & 6  & 118& 20 & 82 & 109& 179      & 52       & 25      \\
MMS+                 & 37  & 17 & 60 & 21 & 22 & 55 & 127      & 4        & 14      \\
SFS+                 & 40  & 6  & 106& 5  & 18 & 7  & 11       & 2        & 6       \\
TCS+                 & 554 & 181& 281& 228& 376& 305& 536      & 140      & 35      \\
CCS+                 & 60  & 57 & 38 & 23 & 35 & 69 & 158      & 30       & 7       \\
HCS+                 & 58  & 118& 73 & 25 & 35 & 81 & 242      & 18       & 25      \\
GCS+                 & 91  & 201& 241& 57 & 108& 132& 246      & 18       & 1      \\
LCS+                 & 91  & 190& 245& 69 & 72 & 174& 327      & 20       & 8      \\
FCS+                 & 13  & 80 & 17 & 9  & 4  & 51 & 93       & 3        & 1      \\
C2C+                 & 89  & 7  & 91 & 24 & 17 & 227& 203      & 14       & 2       \\ 
ACS+                 & 11  & 5  & 20 & 0  & 2  & 19 & 26       & 1        & 1      \\\midrule
SSCS-                & 51  & 139& 40 & 8  & 70 & 118& 168      & 31       & 3       \\ \midrule
\textbf{Total}       &1195 &1007&1330&489 &841 &1397& 2516     & 332      & 127         \\ 
\bottomrule
\end{tabular}
    \label{tab:statistics}
\end{table}

\subsection{Data Annotation} \label{sec:meta-model}

\textbf{Ground-truth Labeling Process.} We used a web tool named \textit{Label Studio}~\cite{label} to show the annotation process. First, we provided annotators with guidelines~\cite{web:code} and conducted three meetings to familiarize them with the conceptual model and the annotation tool. During the annotation process, we published the requirements documents in the tool. For each sentence, the annotators manually labeled the modeling entities. Then, for each pair of entities, the annotators determined whether an interaction existed and labeled its type. The labeled results were used as the ground truth for evaluation. To guarantee the correctness of the labeling results, we built an inspection team consisting of two Ph.D. candidates and four Master students in computer science. All of them are fluent English speakers and have either conducted research on requirements engineering or completed a semester course on requirements engineering. We divided the team into three groups and each group is responsible for four software systems (A, B, or C in Figure \ref{fig:pdd}). The labeled results of one group were reviewed by another group. When a labeled result received different opinions, we hosted a discussion with all team members to decide through voting. In total, we labeled 2,721 requirements from 12 CPSs documents and spent more than 500 person-hours annotating 6,259 entities and 2,975 interactions. Table \ref{tab:statistics} provides a summary of the statistics for \bench{}. In Table \ref{tab:statistics}, ``+'' and ``-'' represent public and private separately. 

\subsection{Quality Evaluation}
\textbf{Consistency Control.} 
To ensure consistency and high quality, we conducted a training phase for all annotators after the three meetings. During this stage, the six annotators were assigned one piece of requirements at a time to complete all annotation tasks. We then calculated the inter-annotator agreement (IAA)~\cite{artstein2017inter} between annotators using Cohen’s Kappa~\cite{mchugh2012interrater}, followed by disagreement discussion and guideline refinement. This process was repeated until the IAA score reached a \textit{substantial agreement} (\ie the IAA score is above 0.6)~\cite{mchugh2012interrater}. Subsequently, the remaining set of requirements was given to the annotators for annotation. The final Cohen's Kappa score for entity labeling is 0.74, and the average Cohen's Kappa for interaction labeling is 0.78. These results demonstrate the quality and reliability of the annotations in \bench{}.

\subsection{\textit{ReqEBench} benchmark}
Figure \ref{fig:pdd} shows a sample from \bench{}. Each sample consists of three components. \textbf{\ding{182} Requirements}: an English text description detailing the requirements of a CPS. \textbf{\ding{183} Entity}: a dictionary that contains all entity concerns and their types. 
\textbf{\ding{184} Interaction}: a dictionary that contains interaction concerns among the entity concerns.

\begin{figure*}
    \centering
    \includegraphics[width=\linewidth]{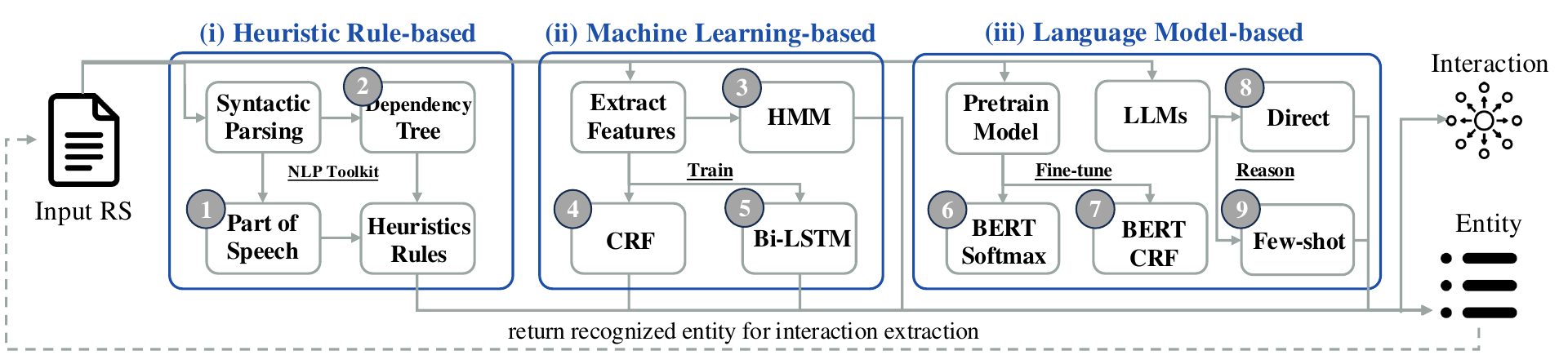}
    \caption{The overview of our comparative study}
    \label{fig:comparative}
\end{figure*}

\section{Comparative Study} \label{sec:comparative_study}
This section begins by defining the formulation of automated requirements concern extraction, followed by a comparative study of three types of alternative solutions. These solutions will be evaluated in Section \ref{sec:resultandanalysis}.

\subsection{Problem Formulation}
Requirements concern extraction involves identifying various dimensions of information (\eg \textit{Physical Device} and \textit{Phenomena Linkage}) from natural language requirements texts. 
As illustrated in Section~\ref{subsec:taxonomy}, 
we focus on two types of extraction tasks, \ie entity recognition and interaction extraction, which can be formulated as follows. 

Let $R = (r_1,r_2,...,r_n)$ denote a sequence of requirements, where each $r_i$ represents a single piece of requirements which is represented by a sentence in natural language. 

\textbf{Entity Recognition.} The entity recognition task is to extract the entity concerns $E = \{(e_i,t_i)\}_{i=1}^{N}$ declared by the taxonomy (\eg Table~\ref{tab:taxo_survey}) from each requirement sentence $r_i$, where $N$ is the number of entities and $e_i$, $t_i$ denote the value and type of the entity concern $i$, respectively. 

\textbf{Interaction Extraction.} Given a requirements sentence $r_i$ and $N$ entities $E = \{(e_i,t_i)\}_{i=1}^{N}$ recognized from the requirements sentence, the interaction extraction task is to predict the interactions concern $I = \{(h_j,n_j,t_j)\}_{j=1}^{M}$ among the recognized entities in $E$ declared in the established taxonomy, where $h_j$ and $t_j$ is the head and tail entity of the $j$-th interaction. $n_j$ is the type of the $j$-th interaction concern.

\subsection{Comparative Solutions}
We consider nine alternative solutions for automating requirements concerns extraction, as shown in Figure \ref{fig:comparative}. Alternatives \circled{1} and \circled{2} are based on heuristic rules; alternatives \circled{3}, \circled{4} and \circled{5} are based on supervised machine learning models (\eg HMM and CRF); alternatives \circled{6}, \circled{7}, \circled{8} and \circled{9} are based on recently popular pre-trained language models. We will elaborate on our alternative solutions in our comparative study next.

\textbf{\textit{(i) Solutions based on heuristic rule.}} We use available NLP toolkits for the syntactic analysis of requirement sentences and apply manually designed heuristic rules for entity recognition and interaction extraction. For solution \circled{1}, we use the SpaCy tool to perform part-of-speech (POS) and apply heuristic rules inspired by previous work~\cite{lucassen2017extracting}. For solution \circled{2}, we employ the StanfordNLP tool to construct a dependency tree (DT) for the requirements and use heuristic rules in the work~\cite{zaki2022rcm}.

\textbf{\textit{(ii) Solutions based on machine learning.}} We explore three machine learning-based solutions: Hidden Markov Models (HMM)~\cite{eddy1996hidden}, Conditional Random Fields (CRF)~\cite{sutton2012introduction}, and Bidirectional Long Short-Term Memory Networks (Bi-LSTM)~\cite{panchendrarajan2018bidirectional}. These solutions are taken from existing NLP and RE literature on information extraction and automated requirements concern extraction. We noticed that the expected input for these solutions should follow a sequence-labeled format. Thus, we first convert the data samples in our \bench{} to BIO format~\cite{sang2003introduction} to train and test these models. 
For solution \circled{3}, the HMM is a classic probabilistic model that can capture sequential features in requirements sentences. It can infer the most likely labels for each word in requirements by estimating state transition and observation probabilities. The input feature dimension for HMM corresponds to the length of the requirements sentence, and the output is the entity concern label for each word. For solution \circled{4}, the CRF is a discriminative probabilistic model that can capture the dependencies between labels of adjacent entities. During our training, we first convert each requirement sentence into a feature vector. The feature vector for each sentence has a dimension of $n \times 6$, where $n$ is the length of the requirements sentence and $6$ is the number of features for each word. For solution \circled{5}, we train a BiLSTM to extract the entity concerns. It can simultaneously utilize both forward and backward contextual information in requirement sentences. Specifically, we first convert the words in the requirements sentences into 20-dimensional word vectors and input the vectors into the BiLSTM. Then, the output from the BiLSTM is mapped to the concern labels space through a fully connected layer. After training, we use the above three solutions to predict the requirements concern label by inputting a new requirements sentence. The three solutions can be used only for the entity concern recognition task.

\textbf{\textit{(iii) Solutions based on language models.}} We employ small language models (\eg BERT) and recently popular LLMs (\eg DeepSeek) to develop solution \circled{6}, \circled{7}, \circled{8}, and \circled{9}. For solution \circled{6} and solution \circled{7}, we fine-tune the BERT on a subset of our \bench{} to expose it to examples of sentences from the RE domains. 
The difference between them is that solution \circled{6} pairs BERT with a softmax classifier, while solution \circled{7} incorporates a CRF layer for the direct classification of the types of concerns. For solution \circled{8} and solution \circled{9}, we construct prompts to guide LLMs to perform entity recognition and interaction extraction without additional fine-tuning. Specifically, solution \circled{8} directly constructs a task description prompt and inputs LLMs to generate the answers. Solution \circled{9} used an advanced technique (\ie few-shot reasoning~\cite{wei2022chain}) to construct the prompt. It will retrieve multiple examples and put them into the prompt. Inspired by previous work~\cite{wang2023gpt}, we use the semantic similarity retrieval strategy to obtain these examples. This involves retrieving examples with close semantics from the training set for each input test requirements sentence. The prompts for solution \circled{8} and \circled{9} can be found in our open source package~\cite{web:code}.

\section{Study Design} \label{sec:studydesign}
To evaluate the performance of the alternative solutions presented in Section \ref{sec:comparative_study}, we conducted a large-scale study to answer three research questions. This section describes the details of our study, including research questions, studied LLMs, metrics, and experiment settings. 

\subsection{Research Questions}
Our study aims to answer the following three research questions (RQs):

\textbf{RQ1: What is the performance of alternative solutions for entity recognition and interaction extraction from requirements?}  We use leave-one-out cross-validation~\cite{dell2023evaluating} to divide \bench{} into training and test datasets. The training dataset is used to train three statistical model-based solutions, fine-tune two BERT-based solutions, and retrieve shots for constructing the LLMs prompts. The performance of them (Figure \ref{fig:comparative}) is evaluated and compared in the test dataset.
 
\textbf{RQ2: What types of errors does the most accurate solution make and their percentage?} Error analysis is essential to guide future improvement. 
In this RQ, we manually review the ground truth and prediction made by the most accurate solution (\ie GPT-4, as shown in Table \ref{tab:rq1_e}) for each test sample in \bench. The error types are summarized, and their percentages are counted. 

\textbf{RQ3: How does the number of shots affect the performance of LLM-based solutions?} Given the context length and response speed limitations of LLMs, we vary the number of shots from 1 to 8. Inspired by a previous study~\cite{wang2023gpt}, we use \textit{Similarity Retrieval} to obtain these shots. The retrieved shots are then used to construct the prompts, which are fed into the LLMs for evaluation.

\subsection{Evaluation Metrics} \label{sec:metrics}
Following previous studies~\cite{shen2023promptner}~\cite{wang2022matchprompt}, we evaluate the effectiveness of entity recognition and interaction extraction using \textbf{precision} ($P$), \textbf{recall} ($R$), and \textbf{F1 score}. Specifically, we first compute the count of correctly recognized entity concerns or interaction concerns ($TP$), the count of entity concerns or interaction concerns recognized by alternative solutions but not present in the gold standard ($FP$), and the count of entity concerns or interaction concerns in the gold standard but not recognized by alternative solutions ($FN$). These values are then aggregated across all entity concerns or interaction concerns and these metrics are computed.


\begin{equation}
\begin{aligned}
    P &= \frac{TP}{TP + FP}, 
    R &= \frac{TP}{TP + FN},
    F1 &= 2 \cdot \frac{P \cdot R}{P + R}
\end{aligned}
\end{equation}

For the entity recognition task, the above metrics (\ie $P$, $R$, and $F1$) are particularly strict. For example, if the ground truth is ``\textit{to monitor and control the temperature of a home}'' and the prediction is ``\textit{to monitor and control a home}'', the prediction would be considered incorrect. However, this prediction may still be reasonable in practical applications. To address this case, the \textbf{rouge-L} metric~\cite{lin2004rouge} is introduced in addition to provide a more comprehensive evaluation. The rouge-L can better reflect the similarity between the predicted modeling entities and the ground truth.

\subsection{Studied LLMs} \label{sec:llms}
Table \ref{tab:llms} shows 5 LLMs for solution \circled{8} and \circled{9} in Figure \ref{fig:comparative}. They are the latest versions of the LLMs released by well-known companies or organizations. They cover closed-source LLMs (\ie gpt-4~\cite{openai2023gpt}) and open-source LLMs (\ie DeepSeek~\cite{guo2025deepseek}, Qwen2.5~\cite{Qwen}, LLama3~\cite{LLama3}, and Gemma~\cite{gemma}). We use their official interfaces or implementations to reproduce these LLMs.

\subsection{Experiment Settings}
The experiment settings of our evaluation are as follows.

\textbf{Dataset Split.} To make sure that any requirements sentence is included in the training dataset or the test dataset, we conduct the \textbf{leave-one-out cross validation} at the system level. 
Specifically, the requirements sentences of one single CPS in \bench{} are retained as the test dataset, and the requirements sentences of the remaining 11 CPSs are included in the training dataset. We repeat this 12 times, since \bench{} contains the requirements sentences of 12 CPSs.

\textbf{Training and Fine-tune Setting.} The ML-based solutions are implemented with Scikit-learn 1.6.1~\cite{pedregosa2011scikit}. Specifically, solution \circled{4} is trained with a maximum of 100 iterations, and solution \circled{5} is trained for 30 epochs using a batch size of 64. The BERT-based solutions are implemented with the Transformers 4.48.3 library~\cite{wolf2020transformers} provided by HuggingFace and operated in Pytorch~\cite{paszke2019pytorch}. Specifically, solution \circled{6} and \circled{7} are based on \textit{bert-based-uncased} model. Detailed experimental details can be found in our replication package~\cite{web:code}.

\textbf{LLMs Setting.} For close-source LLMs, we implement GPT-3.5 and GPT-4 by invoking OpenAI's API~\cite{openaiapi}. For open-source LLMs, we instantiate them with their replication packages and download their pre-trained weight from HuggingFace. The default settings for LLMs are the same, using greedy sampling\cite{chickering2002optimal} with \textit{temperature}=0. We also use the same inference library - vllm~\cite{web:vllm} for LLMs inference to ensure the fairness of our evaluation.

\textbf{Shot Retrieval.} We first use the open-source framework - SimCSE~\cite{gao2021simcse} and the pretrained model - \textit{princeton-nlp/sup-simcse-bert-base-uncased} to compute embeddings for all training examples and the test requirements text. Then, we use faiss~\cite{douze2024faiss} to build the index and get the top 20 similar samples from the training set.

\begin{table}[]
    \centering
    \caption{Evaluated LLMs in our Comparative Study}
    \setlength{\tabcolsep}{5pt}
    \begin{tabular}{ccccc}
\toprule
\textbf{Type}                 & \textbf{Name}   & \textbf{Version}  & \textbf{Context}  & \textbf{Publisher} \\ \midrule
Close-source& GPT-4   & GPT-4-turbo-0409   & 128K  & OpenAI\\ \midrule
\multirow{5}{*}{Open-source}  & DeepSeek-R1    & Distill-Llama-8B           & 128K    & DeepSeek\\
                              & Qwen2.5  & 7B                 & 128K  & Alibaba\\
                              & LLama3.1  & 8B                & 128K    & Meta AI\\
                              & Gemma2   & 7B                 & 8K    & Google\\ \bottomrule
\end{tabular}
    \label{tab:llms}
\end{table}
\section{Result and Analysis} \label{sec:resultandanalysis}

\begin{table}[]
    \centering
    \setlength{\tabcolsep}{1.5pt}
    \caption{Comparison of alternative solutions on entity recognition from CPS requirements}
    \begin{tabular}{ccccccccccccc}
\toprule
\multirow{2}{*}{Solutions} & \multicolumn{2}{c}{SS} & \multicolumn{2}{c}{PD} & \multicolumn{2}{c}{EO} & \multicolumn{2}{c}{ES} & \multicolumn{2}{c}{SR} & \multicolumn{2}{c}{SP} \\ \cmidrule{2-13} 
                           & P/R/F1       & RO     & P/R/F1       & RO     & P/R/F1       & RO     & P/R/F1       & RO     & P/R/F1       & RO     & P/R/F1       & RO     \\ \midrule
\multicolumn{13}{c}{Heuristic rule-based}                                                                                                                                        \\ \midrule
POS                        & 11/9/10      & 34      & 10/8/9       & 28      & 9/11/10      & 33      & 8/16/10      & 33      & 4/8/5        & 21      & 9/15/11      & 41      \\
DT                         & 16/11/13     & 39      & 18/15/16     & 31      & 12/18/14     & 29      & 12/8/10      & 38      & 8/11/9       & 22      & 17/19/18     & 39      \\ \midrule
\multicolumn{13}{c}{Machine learning-based}                                                                                                                                      \\ \midrule
HMM                        & 21/20/19     & 45      & 35/12/14     & 28      & 35/16/20     & 29      & 18/9/11      & 35      & 3/2/2        & 28      & 9/3/5        & 33      \\
CRF                        & 68/9/15      & 13      & 7/3/4        & 6       & 74/10/17     & 15      & 18/3/5       & 19      & 7/8/7        & 6       & 23/2/4       & 13      \\
Bi-LSTM                    & 30/14/19     & 40      & 8/7/7        & 4       & 14/2/3       & 8       & 12/7/9       & 7       & 2/1/1        & 12      & 9/2/3        & 17      \\ \midrule
\multicolumn{13}{c}{Language model-based}                                                                                                                                        \\ \midrule
BERT-SOFT                  & 16/11/13     & 43      & 20/19/19     & 35      & 22/24/23     & 39      & 15/12/18     & 25      & 13/9/11      & 27      & 18/21/19     & 24      \\
BERT-CRF                   & 28/26/27     & 46      & 22/23/22     & 38      & 29/27/28     & 37      & 16/19/17     & 26      & \textbf{16/11/13}& 31      & 17/19/18     & 22      \\
gpt-4 + D& 28/60/38     & 64      & 28/34/29     & 44      & 34/33/33     & 55      & 18/17/16     & 39      & 4/17/6       & 23      & 30/23/25     & 56      \\
Qwen2 + D& 28/52/36     & 63      & 31/43/33     & \textbf{52}& 25/22/23     & 47      & 20/13/15     & 42      & 7/9/8        & 44      & 16/15/15     & 53      \\
LLama3 + D& 26/51/34     & 62      & 14/24/17     & 39      & 22/23/22     & 50      & 22/13/16     & 40      & 6/17/9       & 44      & 28/13/17     & 44      \\
Gemma2 + D& 21/24/22     & 56      & 32/24/24     & 32      & 34/17/21     & 30      & 22/12/14     & 28      & 4/5/4        & 42      & 28/8/12      & 29      \\
DeepSeek + D& 18/45/25     & 49      & 26/32/28     & 47      & 18/23/20     & 47      & 15/5/7       & 29      & 5/9/6        & 44      & 21/7/10      & 38      \\
gpt-4 + F& \textbf{29/64/39}& \textbf{65}& 22/32/25     & 45      & \textbf{34/42/37}& \textbf{59}& 19/21/19     & \textbf{45}& 4/16/6       & 24      & \textbf{32/28/29}& \textbf{54}\\
Qwen2 + F& \textbf{31/55/39}& 63      & \textbf{35/45/36}& 49      & 32/28/29     & 53      & \textbf{29/19/21}& 41      & 6/7/6        & 39      & 18/16/17     & 51      \\
LLama3 + F& 31/55/38     & 62      & 23/29/24     & 42      & 28/35/30     & 57      & 28/17/20     & 44      & 7/15/9       & \textbf{45}& 27/14/17     & 46      \\
Gemma2 + F& 25/39/30     & 64& 26/25/25     & 43      & 32/23/26     & 40      & 16/11/12     & 36      & 4/9/6        & 42      & 26/17/20     & 48      \\
DeepSeek + F& 22/48/29     & 53      & 24/32/26     & 50      & 20/31/24     & 50      & 13/6/8       & 35      & 4/7/5        & 42      & 24/9/12      & 39      \\ \bottomrule
\end{tabular}
    \label{tab:rq1_e}
\end{table}

\begin{table}[]
    \centering
    \caption{Comparison of alternative solutions on interaction extraction from CPS requirements}
    \setlength{\tabcolsep}{2pt}
    \begin{tabular}{ccccccccccccccc}
\toprule
\multirow{2}{*}{} & \multirow{2}{*}{POS}                                & \multirow{2}{*}{DT}                                 & \multirow{2}{*}{\begin{tabular}[c]{@{}c@{}}BERT-\\ SOFT\end{tabular}} & \multirow{2}{*}{\begin{tabular}[c]{@{}c@{}}BERT-\\ CRF\end{tabular}} & \multicolumn{2}{c}{gpt-4}                                                                                 & \multicolumn{2}{c}{Qwen2}                                                                                 & \multicolumn{2}{c}{LLama3}                                                                                & \multicolumn{2}{c}{Gemma2}                                                                                & \multicolumn{2}{c}{DeepSeek}                                                                              \\ \cmidrule{6-15} 
                  &                                                     &                                                     &                                                                       &                                                                      & D                                                   & F                                                   & D                                                   & F                                                   & D                                                   & F                                                   & D                                                   & F                                                   & D                                                   & F                                                   \\ \midrule
IN                & \begin{tabular}[c]{@{}c@{}}18/19\\ /18\end{tabular} & \begin{tabular}[c]{@{}c@{}}21/19\\ /20\end{tabular} & \begin{tabular}[c]{@{}c@{}}21/23\\ /22\end{tabular}                   & \begin{tabular}[c]{@{}c@{}}21/18\\ /19\end{tabular}                  & \begin{tabular}[c]{@{}c@{}}72/59\\ /65\end{tabular} & \begin{tabular}[c]{@{}c@{}}\textbf{74/77}\\ \textbf{/74}\end{tabular}& \begin{tabular}[c]{@{}c@{}}51/51\\ /51\end{tabular} & \begin{tabular}[c]{@{}c@{}}63/60\\ /61\end{tabular} & \begin{tabular}[c]{@{}c@{}}55/57\\ /56\end{tabular} & \begin{tabular}[c]{@{}c@{}}54/58\\ /56\end{tabular} & \begin{tabular}[c]{@{}c@{}}56/55\\ /55\end{tabular} & \begin{tabular}[c]{@{}c@{}}55/62\\ /58\end{tabular} & \begin{tabular}[c]{@{}c@{}}72/55\\ /63\end{tabular} & \begin{tabular}[c]{@{}c@{}}66/52\\ /58\end{tabular} \\
RR                & \begin{tabular}[c]{@{}c@{}}17/25\\ /20\end{tabular} & \begin{tabular}[c]{@{}c@{}}19/12\\ /15\end{tabular} & \begin{tabular}[c]{@{}c@{}}17/25\\ /20\end{tabular}                   & \begin{tabular}[c]{@{}c@{}}24/26\\ /25\end{tabular}                  & \begin{tabular}[c]{@{}c@{}}24/35\\ /29\end{tabular} & \begin{tabular}[c]{@{}c@{}}40/52\\ /45\end{tabular} & \begin{tabular}[c]{@{}c@{}}50/30\\ /38\end{tabular} & \begin{tabular}[c]{@{}c@{}}44/52\\ /48\end{tabular} & \begin{tabular}[c]{@{}c@{}}28/39\\ /33\end{tabular} & \begin{tabular}[c]{@{}c@{}}30/43\\ /36\end{tabular} & \begin{tabular}[c]{@{}c@{}}20/52\\ /29\end{tabular} & \begin{tabular}[c]{@{}c@{}}25/61\\ /35\end{tabular} & \begin{tabular}[c]{@{}c@{}}26/30\\ /28\end{tabular} & \begin{tabular}[c]{@{}c@{}}\textbf{67/35}\\ \textbf{/46}\end{tabular} \\
RC                & \begin{tabular}[c]{@{}c@{}}19/25\\ /22\end{tabular} & \begin{tabular}[c]{@{}c@{}}19/21\\ /20\end{tabular} & \begin{tabular}[c]{@{}c@{}}19/25\\ /22\end{tabular}                   & \begin{tabular}[c]{@{}c@{}}18/24\\ /21\end{tabular}                  & \begin{tabular}[c]{@{}c@{}}24/44\\ /31\end{tabular} & \begin{tabular}[c]{@{}c@{}}\textbf{31/44}\\ \textbf{/36}\end{tabular} & \begin{tabular}[c]{@{}c@{}}15/44\\ /23\end{tabular} & \begin{tabular}[c]{@{}c@{}}19/44\\ /27\end{tabular} & \begin{tabular}[c]{@{}c@{}}27/44\\ /33\end{tabular} & \begin{tabular}[c]{@{}c@{}}27/44\\ /33\end{tabular} & \begin{tabular}[c]{@{}c@{}}8/44\\ /14\end{tabular}  & \begin{tabular}[c]{@{}c@{}}8/44\\ /13\end{tabular}  & \begin{tabular}[c]{@{}c@{}}17/44\\ /25\end{tabular} & \begin{tabular}[c]{@{}c@{}}14/44\\ /22\end{tabular} \\ \bottomrule
\end{tabular}
    \label{tab:rq1_i}
\end{table}

\textbf{RQ1: What is the performance of alternative solutions for entity recognition and interaction extraction from requirements?}

\begin{figure}
    \centering
    \begin{minipage}{0.48\linewidth}
        \centering
        \includegraphics[width=\linewidth]{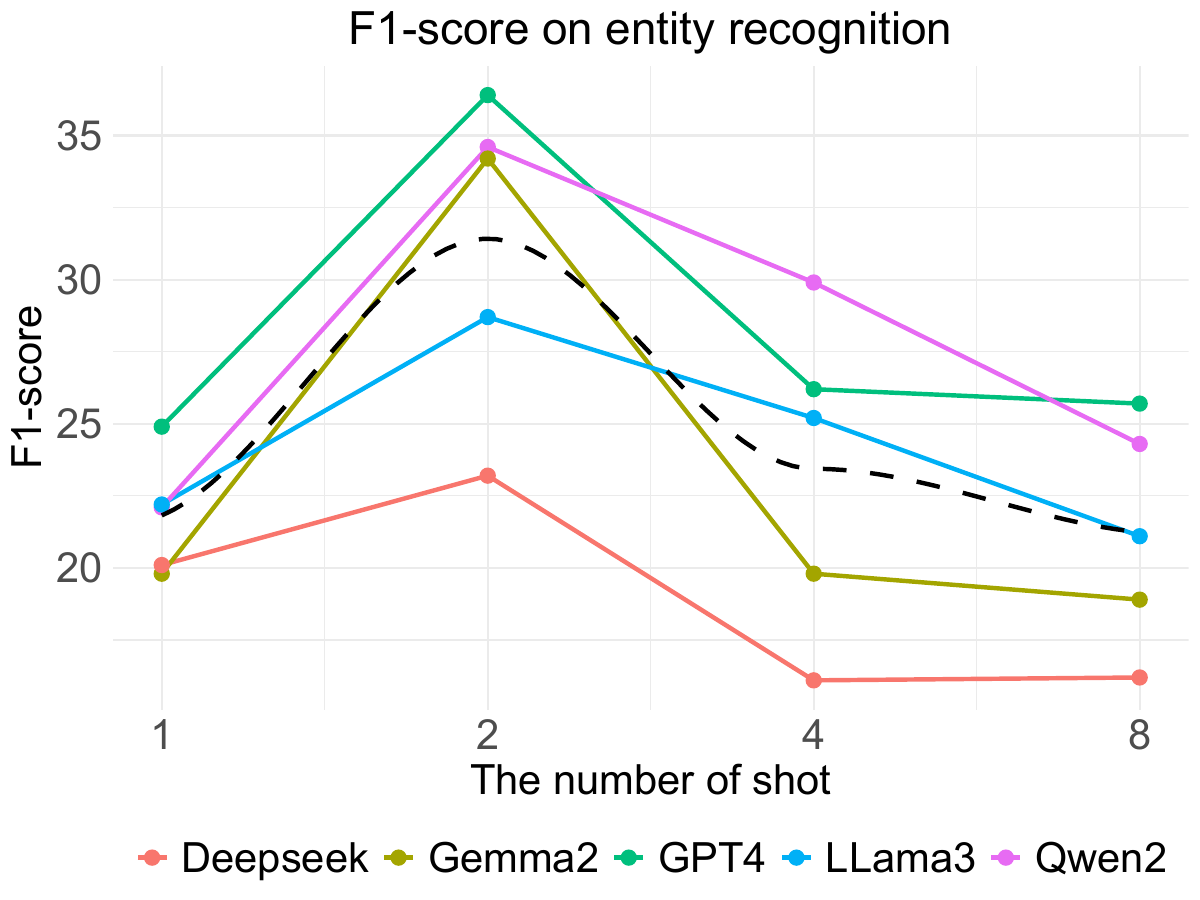}
        \label{fig:rq3-1}
    \end{minipage}
    \hfill
    \begin{minipage}{0.48\linewidth}
        \centering
        \includegraphics[width=\linewidth]{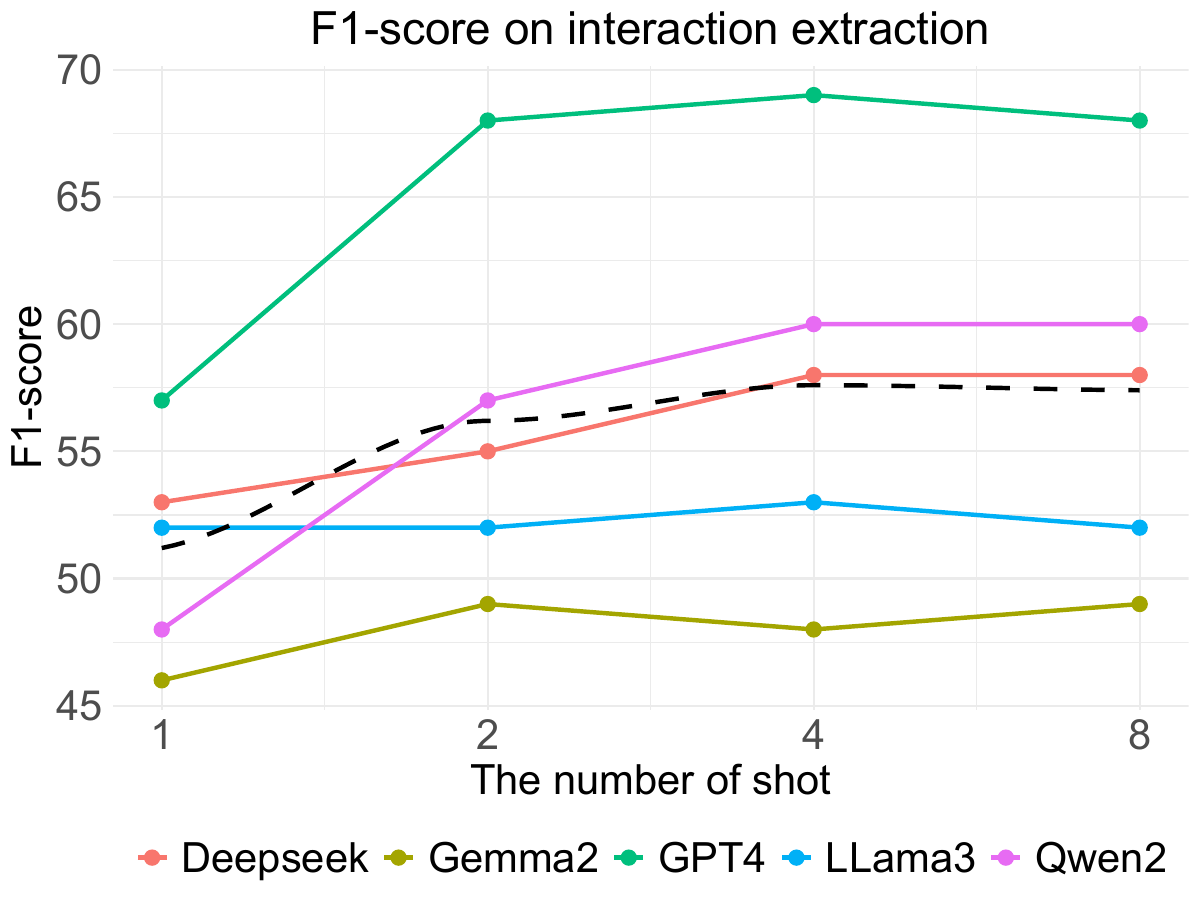}
        \label{fig:rq3-2}
    \end{minipage}
    \caption{The performance comparisons with different shot numbers.}
    \label{fig:rq3}
\end{figure}

\textbf{Setup.} We evaluate the 9 alternative solutions (Figure~\ref{fig:comparative}) using our \bench{} (Section~\ref{sec:benchmark}). For LLM-based solutions, we studied 5 advanced LLMs in table~\ref{tab:llms}. Their prompt is constructed with one shot for solution \circled{9} in this RQ and the performance on more shots is shown in RQ3 (Figure~\ref{fig:rq3}). The evaluation metrics are described in Section~\ref{sec:metrics}, \ie the Precision (P), Recall (R), F1, and rouge-L (ROU). For all metrics, higher scores represent better performance.

\textbf{Results.} Table~\ref{tab:rq1_e} and Table~\ref{tab:rq1_i} show the results of 9 solutions on our \bench{}. In this table, all values are shown as percentages. Each \textbf{bold} represents the best performance for a specific entity or interaction concern. ``D'' and ``F'' represent directly reasoning and few-shot reasoning (\ie 1-shot) separately.

\textbf{Analyses.} 
\textbf{(1) LLM-based solutions significantly outperform traditional methods in requirements concern extraction.} For entity recognition, \textit{GPT-4 + 1-shot} achieves an average F1-score of 0.30, which is more than twice that of the strongest heuristic rule-based method (\ie a F1-score of 0.14 at DT) and over 1.5x that of the best traditional ML method (\ie a F1-score of 0.12 at HMM). In terms of rouge-L, GPT-4 reaches 0.54, compared to only 0.39 for DT and 0.45 for HMM. For interaction extraction, the performance gap is even larger. \textit{GPT-4 + 1-shot} dominates with an F1-score of 0.74 on \textit{``Phenomena Linkage''} interaction, while traditional methods (\eg DT) achieve only 0.20 for the same interaction. 
\textbf{(2) Few-shot reasoning can further enhance the extraction effectiveness of LLM-based solutions.} A consistent improvement is observed when moving from direct prompting (D) to few-shot prompting (F). Specifically, GPT-4 improves from 0.23 (direct) to 0.32 (1-shot) in F1 score and from 0.47 (direct) to 0.49 (1-shot) in rouge-L for entity recognition. These results show that few-shot exemplars consistently yield 10–20\% relative improvement, highlighting the importance of prompt engineering. 
\textbf{(3) Extraction of ``\textit{software requirements}'' entity concern remains particularly challenging.} LLMs consistently underperform on the \textit{``software requirements''} entity concern. Even the best solution (\ie GPT-4 + 1-shot) achieves only 0.10 in F1 score, compared to 0.25–0.33 in F1 for other concerns such as \textit{software system}, \textit{physical device}, and \textit{environment object}. This indicates that ``software requirements'' are more difficult to extract than other entity concerns. We attribute this to their inherently verbose, abstract, and context-dependent descriptions, which span longer textual spans and increase boundary errors. 
\textbf{(4) Current solutions remain insufficient for real-world deployment.} Even the strongest solutions (\ie \textit{GPT-4 + 1-shot}) achieves only 0.32 in recall and 0.25 in F1 on entity concern recognition, meaning that over 60\% of entity concerns are still missed. Similarly, for interaction extraction, while GPT-4 achieves an F1 of 0.74 on the ``\textit{phenomena linkage}'' interaction, its performance drops to 0.36 in F1 on ``\textit{requirements linkage}'' and 0.22 on ``\textit{requirements constraint}'' interactions. These inconsistencies highlight that LLM-based solutions are still far from reliable deployment in industrial-scale requirements engineering.

\begin{boxK}
\small \faIcon{pencil-alt} \textbf{Answer to RQ1:}  LLM-based solutions clearly surpass heuristic and traditional machine learning methods in both entity recognition and interaction extraction, and few-shot prompting further strengthens their effectiveness across model families. However, extracting abstract entities such as ``software requirements'' remains difficult, and overall recall is still limited. This indicates that while LLMs show strong potential for requirements concern extraction, further improvements are necessary for practical deployment.
\end{boxK}

\begin{table*}[]
    \centering
    \setlength{\tabcolsep}{2pt}
    \caption{Analysis of LLMs hallucinations in Entity recognition}
    \begin{tabular}{cccp{7cm}}
    \toprule
    \multicolumn{2}{c}{\bf Error Type} & \bf Type & \bf Case \\
    \midrule
    \multicolumn{2}{c}{\makecell[c]{Type Error \\ (27\%)}} & Input & The Tactical Control System will be capable of being hosted on computers. \\
                                   &       & Ground & Physical Device:[computers] \\
                                   &       & Prediction & Design Domain:[computers] \\
    \cmidrule{1-4}
    \multirow{9}{*}{\makecell[c]{Boundary Error \\ (34\%)}} & \makecell[c]{Contain gold \\ (14\%)} & Input & Tactical Control System provide the capability to control the AV's Identification Friend. \\
                                    &              & Ground & Requirements:[control the AV's Identification Friend] \\
                                    &              & Prediction & Requirements:[provide the capability to control the AV's Identification Friend] \\
    \cmidrule{2-4}
                                    & \makecell[c]{Contained by gold \\ (11\%)} & Input & The thermostat shall allow a user to monitor and control a home’s temperature. \\
                                    &                   & Ground & Requirements: [to monitor and control a home’s temperature] \\
                                    &                   & Prediction & Requirements: [to monitor and control a home] \\
    \cmidrule{2-4}
                                    & \makecell[c]{Overlap with gold \\ (9\%)} & Input & A button providing an opportunity to explore content related to the thematic elements. \\
                                    &                   & Ground & Requirements:[to explore content] \\
                                    &                   & Prediction & Requirements:[explore content related to the thematic elements.] \\
    \cmidrule{1-4}
    \multicolumn{2}{c}{\makecell[c]{Complete Error \\ (21\%)}} & Input & A user shall be able to monitor and control home devices and systems. \\
                                       &       & Ground & Environment Entity:[a user] \\
                                       &       & Prediction & Environment Entity:[home] \\
    \cmidrule{1-4}
    \multicolumn{2}{c}{\makecell[c]{Omitted Entities \\ (18\%)}} & Input & Tactical Control System software provide a windows based graphic operator interface. \\
                                         &       & Ground & Environment Entity:[operator] \\
                                         &       & Prediction & Environment Entity:[] \\
    \bottomrule
    \end{tabular}
    \label{tab:entity_errors}
\end{table*}

\textbf{RQ2: What types of errors does the most accurate solution make and their percentage?}

\textbf{Setup.} To further improve the effectiveness of the automated requirements concern extraction solutions, we perform error analysis for the best current solution (\ie GPT-4). We first use an open source package Gradio~\cite{gradio} to create a web application, which enables better visualization of the requirements, the ground truth, and the prediction results. Then we invited a PhD candidate and a Master student to review these results, summarizing and categorizing the error types.

\textbf{Results.} Table~\ref{tab:entity_errors} and Table~\ref{tab:interaction_errors} show the categorization of the error types and their percentages.

\begin{table*}[]
    \centering
    \setlength{\tabcolsep}{2pt}
    \caption{Analysis of LLMs hallucinations in Interaction extraction}
    \begin{tabular}{ccp{8cm}}
    \toprule
    \bf Error Type & \bf Type & \bf Case \\
    \midrule
    \multirow{3}{*}{\makecell[c]{Type Error \\ (39\%)}} & Input & The CCTNS system should run on multiple browsers. \\
                                & Ground & Requirements Reference:[The CCTNS system, run on multiple browsers] \\
                                & Prediction & Requirements Constraints:[The CCTNS system, run on multiple browsers] \\
    \cmidrule{1-3}
    \multirow{3}{*}{\makecell[c]{Complete Error \\ (31\%)}} & Input & The CMCS system performs limited amounts of real-time data to collect products. \\
                                    & Ground & Requirements Reference:[The CMCS system, to collect products] \\
                                    & Prediction & Requirements Reference:[to collect products, limited amounts of real-time data] \\
    \cmidrule{1-3}
    \multirow{3}{*}{\makecell[c]{Omitted Interactions \\ (30\%)}} & Input & The CCTNS system must be able to export audit trails for specified cases. \\
                                          & Ground & Requirements Reference:[export audit trails, The CCTNS system] \\
                                          & Prediction & Requirements Reference:[] \\
    \bottomrule
    \end{tabular}
    \label{tab:interaction_errors}
\end{table*}

\textbf{Analyses.}  \textbf{(1) There are four types of errors for automated requirements concerns extraction. } Through manual inspection, we summarize errors into four categories: (a) Type Error: an entity concern or interaction concern is correctly extracted but misclassified into the wrong concern category. (b). Complete Error: the extracted entity or interaction is completely different from the gold results. (c) Omitted Error: a correct entity concern or interaction concern is in the omission of the prediction. (d) Boundary Error: a wrong identification of entity concern boundaries, including \textit{contain gold}. {contained by gold}, and \textit{overlap with gold}. The \textit{contain gold} means the predicted entity contains the correct entity. The \textit{contained by gold} means the predicted entity concern is included by the correct entity concern. The \textit{overlap with gold} means the predicted entity concern overlaps with the correct entity concern, but neither contains nor is contained. \textbf{(2) Boundary errors are the most frequent for entity concern recognition.} For entity recognition, boundary errors account for 34\%, which is the main bottleneck in the effectiveness of current automated requirements model extraction. This is because CPS requirements often include long technical expressions (\eg ``audit trails for specified cases'') or multi-word functional descriptions, which challenge LLMs in span detection. In addition, CPS requirements documents contain much terminology and the LLMs do not have sufficient understanding of them.
\textbf{(3) Type errors are the most frequent for interaction concern extraction.}  For interaction extraction, type errors constitute the largest proportion of mistakes (39\%). This means that while LLMs often succeed in identifying that an interaction exists between two entity concerns, they frequently misclassify the type of that interaction. For example, a relation of \textit{Requirement Linkage} may be wrongly labeled as a \textit{Requirement Constraint}. Such errors indicate that the LLM-based solutions capture the surface-level co-occurrence of entities but struggle to differentiate the fine-grained semantics of their relationships. \textbf{(4) Omission is a persistent weakness.} Omitted errors account for roughly one-third of mistakes in interaction extraction (30\%), showing that LLMs often fail to recall less salient or context-dependent entities and relations. For example, requirements that implicitly encode interactions are frequently skipped. This reflects the limited recall of current solutions: LLMs capture prominent concerns but fail to generalize to implicit requirements. Future directions for address these error types are discussed in Section~\ref{sec:future}.

\begin{boxK}
\small \faIcon{pencil-alt} \textbf{Answer to RQ2:}  Error analysis reveals that LLM-based extraction suffers mainly from boundary errors in entity recognition and from type or complete errors in interaction extraction, with omissions also common. These findings suggest that while LLMs can capture many requirement concerns, their precision in classification, boundary detection, and recall still requires substantial improvement to support practical deployment.
\end{boxK}

\textbf{RQ3: How does the number of shots affect the performance of LLM-based solutions?}

\textbf{Setup.} In this RQ, we first retrieve the different numbers of shots and put them in the constructed prompt. Then we feed these prompts to the LLMs and evaluate these LLMs on \bench. Given the context length of LLMs, we set the number of shots to range from 1 to 8. The evaluation metrics are also the precision, recall, and f1 score. 

\textbf{Results.} Figure~\ref{fig:rq3} visual the F1 score of these advanced LLMs (Section~\ref{sec:llms}) on entity concern recognition and interaction concern extraction from 1 to 8 shots, separately.

\textbf{Analyses.} \textbf{(1) Increasing the number of shots generally enhances the effectiveness, but excessive shots degrade their performance.} As the number of shots increases from 1 to 2, the F1 scores of all five studied LLMs shows significant improvement, demonstrating that a few examples can effectively guide the model to better capture the underlying semantics of CPS requirements. However, the performance either stabilizes or decreases when it is beyond 2 shots, indicating that overly long or noisy context may dilute essential cues, increase attention dispersion, and hinder reasoning precision. This trend suggests that while few-shot learning provides valuable contextual grounding, the model’s in-context learning ability remains sensitive to context length and example quality. \textbf{(2) Entity recognition and interaction extraction require different optimal shot ranges.} For entity recognition, all models reach their highest F1-scores with 2-shot prompts, beyond which performance declines sharply. In contrast, interaction extraction tasks benefit from moderate-to-high shot settings (4–6 shots), where additional examples provide richer relational patterns for reasoning. This discrepancy reflects the difference in task complexity: entity recognition mainly relies on local lexical or syntactic cues, whereas interaction extraction depends on relational reasoning that benefits from diversified examples. \textbf{(3) Model families differ in robustness to shot variation.} Among the studied LLMs, GPT-4 and Qwen2 show strong adaptability, maintaining relatively stable performance even when the number of shots increases, while DeepSeek and Gemma2 are more sensitive to contextual noise. This suggests that larger or instruction-tuned models may exhibit higher tolerance to prompt length and better internal calibration mechanisms. These findings imply that future automated requirements concern extraction techniques could adopt adaptive shot selection strategies, \ie dynamically adjusting the number and quality of exemplars according to task type (entity vs. interaction) and model feedback, to balance efficiency and accuracy.


\begin{boxK}
\small \faIcon{pencil-alt} \textbf{Answer to RQ3:}  A small number of high-quality examples (1–2 shots) significantly improve the performance of different LLMs on both entity recognition and interaction extraction, confirming the effectiveness of few-shot prompting in requirements concern extraction. However, excessive shots introduce noise and degrade performance, showing that models are sensitive to context length and example quality. 
\end{boxK}

\section{Discussion} \label{sec:discussion}

\subsection{Threats to validity} 
\textbf{Construct Validity} concerns the relationship between treatment and outcome. The threat comes from the rationality of the research questions we posed. We aim to conduct a comparative study on automated requirements concern extraction solutions. To achieve this goal, we focus on benchmark construction, empirical evaluation, error analysis, and the impact of shot numbers. We believe that these questions have great potential to provide insights for the subsequent development of automated requirements concern extraction.

\textbf{Internal Validity} concerns the threats to the way we carry out our study. The first threat is related to the construction of our \bench{} (Section \ref{sec:benchmark}). To construct the benchmark, we manually annotated the requirements documents. We acknowledge that these annotations are somewhat subjective. To mitigate this threat, we provided the annotators with an annotation guide and held three meetings to learn about the taxonomy of requirements concerns and the annotation tool. Then each annotator independently annotated the benchmark and each label was cross-validated. In addition, we also calculated the consistency scores between the annotators. The second threat relates to the setup of LLMs when addressing RQ1 and RQ3. LLMs may show different performances under different decode strategies or inference frameworks. To mitigate this threat, we set LLMs as greedy decoding and the same inference framework. 


\textbf{External Validity} concerns the threats to generalize our findings. 
The first threat is the representativeness of our benchmark. To mitigate this threat, the requirements documents in our benchmark include various types of CPSs, covering a wide range of application domains (\eg transportation, military, and aerospace). This ensures that the collected requirements documents can represent the diversity of CPS requirements. The second threat is the selection of LLMs. We select the latest version of the LLMs with around 7 billion parameters released by well-known companies or organizations. This is to ensure the practicality and efficiency of LLMs in real-world applications, as LLMs with excessive parameters might not be feasible for organizations to deal with requirements tasks due to the limitation of computing resources.

\subsection{Future Directions}  \label{sec:future}

Current LLMs have shown remarkable potential in automating requirements concern extraction. However, our empirical study also reveals persistent challenges such as boundary ambiguity, incomplete recognition, and conceptual confusion between concern types. To advance this line of research, several key directions can be pursued. 

\textbf{1. Enhance domain terminology understanding.} Boundary errors often arise from LLMs' insufficient understanding of domain-specific terminology. Future research should explore ways to integrate CPS-related knowledge into LLMs through retrieval-augmented generation, continued pretraining, or domain-specific instruction tuning. For example, integrating structured device ontologies, component taxonomies, or standardized engineering vocabularies into prompts could improve precision in identifying technical entities. Similarly, exposing LLMs to authentic CPS documentation, such as control system manuals, aerospace specifications, or embedded software design reports, could enhance their contextual comprehension and reduce semantic drift in recognition tasks.

\textbf{2. Strengthen knowledge of key concerns related to CPS requirements.} LLMs still struggle with specialized concerns such as ``\textit{software requirements}'' and ``\textit{physical devices}'', which require understanding both linguistic and structural relationships. To address this, future work can combine concept-aware prompting with rule-based or symbolic post-processing to maintain concern consistency. Instruction-tuning on structured concept datasets would further enable LLMs to internalize the compositional semantics of requirements concern concepts, bridging the gap between linguistic fluency and formal structure understanding.

\textbf{3. Integrate knowledge from multiple LLMs.} Different LLMs demonstrate distinct strengths, \ie some excel at reasoning and abstraction, while others specialize in factual recall or linguistic precision. A hybrid or collaborative model design, where domain-tuned smaller models act as knowledge retrievers or verifiers, and general-purpose LLMs handle reasoning and abstraction, could combine their advantages. Such multi-model collaboration has the potential to balance accuracy, interpretability, and scalability in automated requirements extraction.

\textbf{4. Human-in-the-loop refinement and adaptive learning.} Despite advances in automation, human expertise remains indispensable to interpret incomplete or ambiguous requirements. A promising direction is to develop interactive tools that allow analysts to visualize, verify, and correct model output. These human corrections can be recorded and reused for targeted retraining, focusing on frequent error categories such as boundary and type errors. Over time, such human-in-the-loop pipelines would support continuous improvement and build trust in the use of LLMs for critical engineering tasks.


\section{Related works}

\subsection{Solutions on Automated Requirements Concern Extraction}
Automated requirements concern extraction aims to extract key concerns from unstructured natural language requirements documents. Existing solutions can be divided into three categories, namely heuristic rule-based, machine learning-based, and language model-based solutions.  

\textbf{Heuristic rule-based Solutions.} Heuristics rule-based solutions typically employ natural language processing (NLP) techniques to parse requirements texts and develop linguistic rules to extract target requirements concerns from the parsed data. Güneş et al.~\cite{gunecs2020automated} and Lucassen et al.~\cite{lucassen2017extracting} used the Spacy tool to parse part-of-speech (POS) tags of use stories and developed heuristic rules to extract concerns in goal models and process models, separately. Herchi et al.~\cite{herchi2012user} applied NLP techniques for POS tagging and syntactic parsing of requirements texts, developing 12 heuristic rules to extract concerns in class diagrams. Casagrande et al.~\cite{casagrande2014nlp} and Zaki‐Ismail et al.~\cite{zaki2022rcm} employed the Stanford NLP toolkit to parse the semantic structure of requirements texts and use rule-based algorithms to extract concerns in semi-formal representation models. 

\textbf{Machine learning-based Solutions.} Machine learning-based solutions emphasize the use of statistical models and simple neural networks to extract requirements concerns. These solutions typically involve training classifiers on annotated datasets to extract requirements concerns. Unlike heuristic rules-based solutions, machine learning-based solutions learn from annotated data and can adapt to different requirements texts. Saini et al.~\cite{saini2022machine} provided an incremental learning strategy empowered by machine learning to extract domain models by analyzing practitioner decisions. Yang et al.~\cite{yang2022towards} trained a classifier to tag each requirement sentence as describing a class or relation. Bencomo et al.~\cite{bencomo2019ram} proposed a POMDP-based solution to extract and update concerns in runtime models using Bayesian inference and adaptation mechanisms. Burgueño et al.~\cite{burgueno2019lstm} proposed an LSTM-based solution to infer model transformations from sets of input-output model pairs.

\textbf{Language model-based Solutions.} Recent years have witnessed the emergence of language models, \eg Transformers~\cite{vaswani2017attention} and ChatGPT~\cite{gpt-3.5}. Language model-based solutions can be divided into two categories. \textit{Training small language models.} This category involves fine-tuning pre-trained small language models (\eg BERT~\cite{devlin2018bert}) on human-labeled datasets. It formulates the task into structured information extraction tasks and adopts the fine-tuned models to extract requirements concerns, \eg actor and action. For example, Jin et al.~\cite{jin2023automating} created four datasets and fine-tuned BERT using them to extract concerns in problem diagrams. Zhou et al.~\cite{zhou2022assisting} fine-tuned BERT and combined logical reasoning to extract requirements concerns in goal models. \textit{Reasoning large language models.} Reasoning LLMs (\eg ChatGPT or GPT-4) for automated requirements concern extraction often involves carefully designing prompts. The key advantage is that analysts can directly use LLMs to extract requirements concerns, eliminating the need for training on labeled datasets. Chen et al.~\cite{chen2023automated} conducted experiments for GPT-3.5 and GPT-4 with various prompts to evaluate the extraction of concerns in domain models. 



In this paper, we conduct a comparative study of these three types of solutions, hoping to reveal their performance in real-world CPSs through our \bench.

\subsection{Requirements Concern Extraction Benchmarks}
Benchmarks have become a mainstream method for evaluating the effectiveness of proposed approaches. They enable researchers to compare the performance of various methods. However, to the best of our knowledge, no publicly available requirements concern extraction benchmark aligns with real-world CPSs. Most existing studies rely on a few small cases to evaluate the performance of their solutions. For example, Ruan et al.~\cite{ruan2023requirements} use a digital home system to illustrate the effectiveness of their framework. Camara et al.~\cite{camara2023assessment} use ten small cases (\ie about three to six classes) to evaluate ChatGPT's ability. 
Arora et al.~\cite{arora2016extracting} use four cases to assess the performance of their approach in extracting domain models. The four cases contain 110 to 380 pieces of requirements. In addition, some researchers built specific benchmarks that were tailored to their concerns. For example, Zhou et al.~\cite{zhou2022assisting} constructed a dataset for goal model extraction.

Compared with them, our \bench{} is a new public requirements concern extraction benchmark aligned with real-world CPSs across multiple dimensions.

\section{Conclusion} \label{sec:conclusion}
This paper proposes a new benchmark for the extraction of the real world requirements model named \bench{}. It aligns with real-world CPS requirements in multiple dimensions (\eg scale and complexity) and reduces the gap between evaluation and practical application. We conducted a comparative study on three types of automated requirements extraction solutions using \bench{}. The results reveal their strengths and weaknesses in real-world CPS requirements. We also summarize the error type and provide insight for future directions by analyzing failed cases. We hope \bench{} and our empirical findings can facilitate the evaluation and development of automated requirements concern extraction.



\bibliographystyle{ACM-Reference-Format}
\bibliography{myref}

\end{document}